\documentclass[11pt,amssymb,nofootinbib]{revtex4}
\usepackage[latin9]{inputenc}
\usepackage{textcomp}
\usepackage{amsmath}
\usepackage{amssymb}
\usepackage{esint}
\usepackage{graphicx}
\usepackage{epstopdf}
\usepackage[english]{babel}

\makeatletter
\@ifundefined{textcolor}{}
{%
 \definecolor{BLACK}{gray}{0}
 \definecolor{WHITE}{gray}{1}
 \definecolor{RED}{rgb}{1,0,0}
 \definecolor{GREEN}{rgb}{0,1,0}
 \definecolor{BLUE}{rgb}{0,0,1}
 \definecolor{CYAN}{cmyk}{1,0,0,0}
 \definecolor{MAGENTA}{cmyk}{0,1,0,0}
 \definecolor{YELLOW}{cmyk}{0,0,1,0}
}

\usepackage{amsthm}\usepackage{amscd}\usepackage{bm}\usepackage{bbm}

\let\vec\boldvec%

\makeatother

\begin{document}

\title{Spontaneous photon emission from a non-relativistic free charged particle in collapse models: A case study}


\author{Angelo Bassi}

\email{bassi@ts.infn.it}

\affiliation{Department of Physics, University of Trieste, Strada Costiera 11,
34151 Trieste, Italy}

\affiliation{Istituto Nazionale di Fisica Nucleare, Trieste Section, Via Valerio
2, 34127 Trieste, Italy}

\author{Sandro Donadi}

\email{sandro.donadi@ts.infn.it}

\affiliation{Department of Physics, University of Trieste, Strada Costiera 11,
34151 Trieste, Italy}

\affiliation{Istituto Nazionale di Fisica Nucleare, Trieste Section, Via Valerio
2, 34127 Trieste, Italy}


\begin{abstract}
We study the photon emission rate of a non relativistic charged particle interacting with an external classical noise through its position. Both the particle and the electromagnetic field are quantized. Under only the dipole approximation, the equations of motion can be solved  exactly for a free particle, or a particle bounded by an harmonic potential. The physical quantity we will be interested in is the spectrum of the radiation emitted by the particle, due to the interaction with the noise. We will highlight several properties of the spectrum and clarify some issues appeared in the literature, regarding the exact mathematical formula of a spectrum for a free particle.
\end{abstract}

\maketitle
\section{Introduction}
The dynamics of charged particles has always been at the heart of both theoretical and experimental physics, both in its classical and quantum version. Electromagnetism is the best known interaction experimentally, and from the theoretical point of view it still puzzles because of the $1/r$ dependence and the associated divergences. 

In this paper we analyze the dynamics of a non-relativistic charged quantum particle (either free or bounded by an harmonic potential) under the influence of an external classical noise coupled to the particle's position. We will focus on the spectrum of the radiation emitted by the particle, due to the interaction with the noise. Under only the dipole approximation, the equations can be solved exactly, and important counterintuitive features of the model appear clearly, which would be hidden in a perturbative analysis. 

The behavior of charged particles is particularly important in the contest of collapse models, because up to now it sets the strongest bound on the possible range of values, which the phenomenological parameters defining these models can take~\cite{sci,ad1}.  The idea is the following. Collapse models modify the standard quantum evolution given by the Schr\"odinger equation, by adding nonlinear and stochastic terms which induce the collapse of the wave function. By modifying the dynamics, they modify the physical predictions, therefore they can be tested against standard quantum theory.  By comparing the theoretical predictions with available experimental data, one can set a bound on the collapse parameters. Indeed, nowadays there is a lot interest and  experimental effort in this direction~\cite{mm0,mm1,mm2,mm3,mm4}. 

For the above mentioned reasons, research on the spontaneous emission of radiation from charged particles within collapse models has been quite active. Preliminary results were found by P. Pearle and collaborators~\cite{fu0,fu1,fu2} using the GRW model~\cite{grw,prep} and by L. Di\'{o}si and B. Luk\'{a}cs~\cite{diosi} using models where wave-like stochastic gravitational disturbances destroy macroscopic coherence. The first theoretical calculation using the CSL model~\cite{prep,www,csl} (for many reasons, the most physical collapse model) has been carried out by Q. Fu~\cite{fu3}, to the first meaningful perturbative  order, and relative to the free particle case. The calculation has been confirmed by S.L. Adler and F.M. Ramazanoglu~\cite{ar}, and generalized to hydrogenic atoms and to non-white noises. More recently, one of us (A.B.) and D. D\"urr have done a similar calculation~\cite{bd} using the QMUPL model~\cite{qmupl}. This model is simpler from the mathematical point of view, and allows for an exact analytical treatment of the problem; moreover, in an appropriate limit it should reproduce the same results as those of the CSL model, as argued in~\cite{bd}. However, the free particle's photon-emission rate turns out to be twice that predicted by the CSL model. 

This discrepancy, which initially seemed of little importance, revealed many subtleties in the implementation of standard quantum field perturbative methods in the context of stochastic models; in particular, it revealed that for colored noises, terms appear in the radiation emission formula, which look unphysical~\cite{abd}. As  explained at the beginning of~\cite{abd}, the authors of~\cite{fu3,ar} obtained the correct result by neglecting terms which in principle should have not been neglected (see Eq. (45) of~\cite{abd} and subsequent discussion). The authors of~\cite{bd} first spotted out such a mistake. Ref.~\cite{abd} showed that, in spite that such terms should be taken into account, they are not physically relevant. Here we wish to come back on this issue, and further clarify the matter. 

After having clarified the origin of the unphysical extra term, in the last section we perform the computation of the emission rate using a semiclassical method, following the approach of~\cite{ad1,diosi}. We show that, in such a case, the computation is somehow easier and the unphysical term does not appear.

\section{The model}
The setup is the same as in~\cite{bd} and we summarized it in the supplementary material to this letter~\cite{supp}. We consider a non-relativistic charged particle bounded by an harmonic potential, interacting with a vector noise ${\bf w}(t)$ according to the following (stochastic) Shr\"odionger equation:
\begin{equation} \label{eq:qmupl}
\frac{d}{dt} \psi_t = \!\left[ - \frac{i}{\hbar} H + i
\sqrt{\lambda} {\bf q} \cdot {\bf w}_t \right] \!\psi_t,
\end{equation}
where ${\bf q}$ is the position of the particle, and $\lambda$ a positive coupling constant. The electromagnetic field is also quantized and the full quantum Hamiltonian $H$ reads:
\begin{equation} \label{eq:H}
H \; = \; \frac{1}{2m_0}({\bf p} - e {\bf A})^2 + \frac{1}{2}
\kappa\, {\bf q}^2 + \frac{1}{2}\epsilon_0 \int d^3x \, \left[ {\bf
E}^2 + c^2 {\bf B}^2 \right],
\end{equation}
where $m_0$ is the bare mass of the particle, $\kappa$ is the force constant of
the harmonic term, ${\bf A}$ is the vector potential, ${\bf E}$ and ${\bf B}$
the electric and magnetic fields respectively, $e$ is the electric charge, $c$
the speed of light and $\epsilon_0$ the vacuum permittivity.  In the calculations, the gauge ${\bf \nabla} \cdot {\bf A} = 0$ and $V = 0$ is used, where
$V$ is the electromagnetic scalar potential. Under the dipole approximation, the equations of motions have been exactly solved~\cite{bd,supp} and the dynamical evolution of all physical quantities can be unfold. 

As anticipated, we focus our attention on the emission rate, defined as follows:
\begin{equation} 
\frac{d}{dk} \Gamma_{k}(t) =8\pi k^{2} \frac{d}{dt} {\mathbb E} [\langle \phi | a_{\bf k \mu}^{\dagger}(t) a_{\bf k \mu}(t) | \phi \rangle]
\end{equation}
where $|\phi\rangle$ is the initial state and $a_{\bf k \mu}^{\dagger}(t)$, $a_{\bf k \mu}(t)$ are the creation and annihilation operators of a photon of momentum ${\bf k}$ and polarization $\mu$, in the Heisenberg picture. The symbol ${\mathbb E} [ \cdot ]$ denotes the stochastic average. The factor $8\pi k^{2}$ arises because a sum over the polarization states and directions of the photon momentum is taken. In what follows, we take $|\phi\rangle := |\psi\rangle \otimes |\Omega\rangle$, where $|\psi\rangle$ is the initial state of the particle and
$|\Omega\rangle$ is the vacuum state for the electromagnetic field. From a simple inspection to the equations for $a_{\bf k \mu}^{\dagger}(t)$ and $a_{\bf k \mu}(t)$ (see Eqs. (22) and (23) of~\cite{bd}; they have been reported in Eqs. (10) and (11) of~\cite{supp}) one can easily see that:
\begin{eqnarray} \label{eq:dfgsdtua}
\frac{d}{dk}\Gamma_{k}(t) & = & \frac{\lambda\hbar e^{2}}{2\pi^{2}\varepsilon_{0}c^{2}}\omega_{k}\frac{d}{dt}\int_{0}^{t}ds\int_{0}^{t}ds'G_{1}^{-}\left(k,t-s\right) \nonumber \\
& & \cdot {G}_{1}^{+}\left(k,t-s'\right)f\left(s-s'\right).
\end{eqnarray}
Here, $\omega_k = c k$ ($k = | {\bf k} |$) is the frequency
corresponding to the wave vector ${\bf k}$ and $f\left(s-s'\right)$ is the correlation function of the noise, i.e. we are assuming:
\begin{equation}
\mathbb{E} [ w_{i}(s) w_{j} (s')] = \delta_{ij} f(s-s'),
\end{equation}
while its average is zero. 
The functions ${G}_{1}^{+}$ and ${G}_{1}^{-}$ are defined as follows:
\begin{equation} \label{eq:fdgdf}
G^{\pm}_1(k,t) \; = \; \int_{\Gamma} \frac{dz}{2\pi i} \, \frac{z
e^{zt}}{(z \pm i \omega_k) H(z)}; 
\end{equation}
the contour $\Gamma$ refers to a
line parallel to the imaginary axis, lying to the right of all singularities of
the integrand.
The function $H(z)$ depends on the particle's form factor, and in case of a point-like particle it diverges. One can set up a standard renormalization procedure in order to cure the divergence. After renormalization, $H(z)$ takes the form:
\begin{equation} \label{eq:Hfun}
H(z) = \kappa + z^2 \left[ m - \beta z \right],
\; \beta = \frac{e^2}{6 \pi \epsilon_0 c^3} \simeq 5.71 \times 10^{-54} \text{Kg s},
\end{equation}
where $m$ is the renormalized mass (different from $m_0$). The constant $\beta$ is the coefficient in front of the Abraham-Lorentz
force, which is responsible for the runaway behavior of the corresponding
Abraham-Lorentz equation. Now we have all the ingredients to compute the emission rate both in the free particle case and in that of an harmonic oscillator.

\section{Free particle}
When the particle is free ($\kappa = 0$), the integral in Eq.~\eqref{eq:fdgdf} is easy to solve and the functions $G^{\pm}_1(k,t)$ take the form~\cite{com2}:
\begin{equation}
G_{1}^{\pm}\left(k,t\right)=\mp\frac{i}{m\omega_{k}}\pm\frac{ie^{\mp i\omega_{k}t}}{\omega_{k}\left(m\pm i\beta\omega_{k}\right)}+\frac{e^{mt/\beta}}{\left(m/\beta\right)\left[m\pm i\beta\omega_{k}\right]}.
\end{equation}
The last term gives rise to the runaway behavior, which is expected after the renormalization procedure; we pragmatically dismiss it by ignoring its contribution. The second term oscillates in time, while the first term is constant. When computing the product between $G_{1}^{-}$ and $G_{1}^{+}$ terms arise which oscillate with a rate much bigger than typical experimental times, and thus average to zero. Accordingly, one can neglect all these oscillatory contributions and keep only the constant terms. 

 In the large-time limit ($t \rightarrow \infty$), the rate takes the form:
\begin{equation} \label{eq:pfctu}
\frac{d}{dk}\Gamma_k= \frac{1}{2} \left. \frac{d}{dk}\Gamma_k \right|_{\text{\tiny white}} \cdot \left[\tilde{f}\left(0\right)+\frac{m^2}{m^{2}+\beta^{2}\omega_{k}^{2}}\tilde{f}\left(\omega_{k}\right)\right],
\end{equation}
where: 
\begin{equation}\label{eq:ftilde}
\tilde{f}(k) = \int_{- \infty}^{+ \infty} f(s) e^{iks} ds
\end{equation}
is the Fourier transform of the correlation function of the noise, and:
\begin{equation} \label{eq:udsfdt}
\left. \frac{d}{dk}\Gamma_k \right|_{\text{\tiny white}} =  \frac{\lambda\hbar e^{2}}{\pi^{2}\varepsilon_{0}c^2 m^2  \omega_k}.
\end{equation}
is the rate in the case of a white noise field ($f(s-s') = \delta(s-s')$), to first perturbative order ($\beta = 0$), as we shall now see. We focus on this quantity because this is the one around which the discussion was centered, in the literature. 
If we keep only the leading order terms in the electric charge $e$, i.e. if we set $\beta = 0$, the emission rate becomes:
\begin{equation} \label{eq:sdfsdl}
\frac{d}{dk}\Gamma_k \; = \; \frac{1}{2} \left. \frac{d}{dk}\Gamma_k \right|_{\text{\tiny white}} \cdot [\tilde{f}(0) + \tilde{f}(\omega_k) ]. 
\end{equation}
Note that the first order's approximation in the electromagnetic constant $e$ is good so long as $\omega_k \ll m/\beta$, which means a photon's energy $E_k \ll 6.62 \times 10^{5}$keV, when the emitting particle is an electron. All available experimental data~\cite{fu3} fall well within this range. Moreover photons that are not in this range have energies larger by more than three order of magnitude than the electron mass: since our computation is non relativistic, this situation is implicitly discarded. Therefore the perturbative approach is well justified in this case, except for the problems we will now discuss.

In the {\it white noise} limit ($\tilde{f}(k) = 1$), the exact formula~\eqref{eq:pfctu} for the emission rate necessarily coincides with Eq.~(47) of Bassi and D\"urr~\cite{bd}, while the leading contribution with respect to the electric charge $e$ (Eq.~\eqref{eq:sdfsdl} with $\tilde{f}(k) = 1$) is twice the perturbative result of Adler and Ramazanoglu~\cite{ar}. This can seem of little importance. However in the {\it non-white noise} case, the result is rather surprising. The photon emission rate depends not only on the value of the Fourier transform $\tilde{f}(k)$ of the noise correlator at $k=\omega_k$, the frequency of the emitted photon, as the standard energy conservation argument would suggest. It depends also on its value at zero~\cite{abd}. This is a rather counterintuitive and somehow unphysical result. Moreover, it makes a big difference concerning its implications for collapse models. While, without the $\tilde{f}(0)$ term, the magnitude of the rate crucially depends on the shape of the time correlator, and in particular on whether or not there is a cut--off at some (high) frequency, the presence of this extra term nullifies such a dependence (a cut-off at very low frequency is less likely, from the physical point of view). This, in turn, directly affects the possible upper bounds on the collapse parameters, which are set by experimental constraints~\cite{ar}.

From the mathematical point of view, there is no way out from this result, as our calculations are exact, apart from the dipole approximation. A different insight can be gained by analyzing the case of a particle bounded by an harmonic potential.

\section{Harmonic oscillator}
The different behavior of an harmonic oscillator, mathematically speaking, lies in the fact that the function $H(z)$ defined in Eq.~(\ref{eq:Hfun}) has a different dependence on $z$ compared to a free particle. The functions $G^{\pm}_1(k,t)$ entering Eq.~(\ref{eq:dfgsdtua}) change, and become:
\begin{equation}
G^{\pm}_1(k,t) = \sum_{l=1}^{3}\frac{z_{l}e^{z_{l}t}}{\left(z_{l}\pm i\omega_{k}\right)}\left[\frac{z-z_{l}}{H\left(z\right)}\right]_{z=z_{l}}\mp\frac{i\omega_{k}e^{\mp i\omega_{k}t}}{H\left(\mp i\omega_{k}\right)},
\end{equation}
where $z_{\ell}$ are the three zeros of $H(z)$. For a more explicit form of these functions see~\cite{supp}. Their leading order expression with respect to the frequency $\omega_0 = \sqrt{\kappa/m}$ of the oscillator (in what follows, we will be interested in the free particle limit: $\omega_0 \rightarrow 0$) is:
\begin{equation} \label{eq:sol}
z_1  =  \frac{m}{\beta} +  o(\omega_0), \qquad
z_{2,3} = - \frac{\omega_0^2 \beta}{2m} \pm  i
\omega_0 + o(\omega_0^3),
\end{equation}
The term with $z_1$ gives the usual runaway contribution, therefore we dismiss it. 
The emission rate now takes a rather complicated expression, which is reported in Eq.~(23) of~\cite{supp}. As usual, there are oscillatory terms which rapidly fluctuate and therefore can be neglected. The major difference from the free particle case is the presence of exponentially damped terms, originating from the solutions $z_2$ and $z_3$. These exponential contributions become 1 in the free particle limit ($\omega_0 = 0$), but vanish asymptotically in the case of an harmonic oscillator. Moreover they vanish very rapidly: In fact, taking $\omega_0 = E/\hbar \simeq 3.29 \times 10^{15}$Hz, where we have taken $E = 13.6$ eV the energy of the ground state of the hydrogen atom, one has $\omega_0^2 \beta / 2m \simeq 3.39 \times 10^{7} \text{s}^{-1}$, while typical experiments last years~\cite{fu3}.

Coming back to the harmonic oscillator, by neglecting all runaway, oscillatory and exponentially vanishing terms, we are left with the following expression:
\begin{equation}\label{eq:harmonicoscil}
\frac{d}{dk}\Gamma_k = \frac{\lambda\hbar e^{2}}{2\pi^{2}\varepsilon_{0}c^2} \frac{\omega_{k}^3}{m^{2}\left(\omega_{0}^{2}-\omega_{k}^{2}\right)^{2}+\beta^{2}\omega_{k}^{6}}\tilde{f}\left(\omega_{k}\right).
\end{equation}
This formula---in the free particle limit---reduces to:
\begin{equation} \label{eq:fdgdfg}
\frac{d}{dk}\Gamma_k = \frac{1}{2} \left. \frac{d}{dk}\Gamma_k \right|_{\text{\tiny white}} \cdot \frac{m^2}{m^{2}+\beta^{2}\omega_{k}^{2}}\tilde{f}\left(\omega_{k}\right).
\end{equation}
When compared with Eq.~(\ref{eq:pfctu}), we see that the ``unphysical'' term associated to $\tilde{f}\left(0\right)$ has disappeared, and with it also the factor of 2 difference with respect to the CSL perturbative calculation of Fu~\cite{fu3} ($\beta = 0, \tilde{f}\left(\omega_{k}\right) = 1$) and of Adler et all~\cite{ar,abd} ($\beta = 0, \tilde{f}\left(\omega_{k}\right)$ generic).

Since in our case the whole calculation has been done non perturbatively, the mathematical reason behind such a difference is easy to find: the large time limit and the free particle limit do not commute. The reason for this, as previously noted, is that the free particle limit $\omega_0 \rightarrow 0$ implies also $z_{2,3} \rightarrow 0$. Accordingly, if one takes the large time limit before the free particle limit, then the terms in the formula for the emission rate depending on $z_{2,3}$ vanish exponentially and do not contribute to the asymptotic rate. On the other hand, if one takes the free particle limit before the large time limit, then those terms do not vanish exponentially anymore, and do contribute to the final rate, with a term proportional to $\tilde{f}\left(0\right)$. 

With the mathematical clarification comes also the physical explanation. In doing the free particle limit before the large time limit, one is assuming that the particle is perfectly free at all times, forever. Vice-versa, by computing the large time limit before the free particle limit, one is assuming that the particle feels---sooner or later---the edges of a (harmonic) potential. Therefore the two different orders of limits imply two different physical situations, hence two different behaviors.

Since particles, which are forever free, do not exist in nature, we come to the following conclusion. Eq.~(\ref{eq:pfctu}), and the result of Bassi and D\"urr~\cite{bd}, is mathematically correct for a perfectly free particle, with the unphysical term included. However, at the practical level such a term is irrelevant, since real particles are never truly free. Then, the physically relevant formula is that of Eq.~(\ref{eq:fdgdfg}), without the unphysical term: this equation, in the white noise limit, and to first perturbative order ($\beta = 0$), coincides with the CSL calculation of Fu~\cite{fu3} and of Adler et al.~\cite{ar,abd}.

\section{First order perturbation analysis}
A deeper insight on the problem can be obtained by considering a perturbative approach with respect to the electromagnetic coupling constant $e$. The first-order analysis is equivalent to taking first the limit $\beta \rightarrow 0$, and then the limit $t \rightarrow \infty$ in the formula for the emission rate, Eq.~\eqref{eq:dfgsdtua}. For a free particle, the final result, i.e. Eq.~\eqref{eq:sdfsdl} with the unphysical term included, does not change. Therefore, in this case the exact and perturbative calculations match.  However, in the case of an harmonic oscillator the perturbative result is completely different~\cite{supp}:
\begin{equation} \label{eq:bzero}
\frac{d}{dk}\Gamma_k = \frac{e^{2}\hbar\lambda}{4\pi^{2}\varepsilon_{0}c^{2}m^{2}}\, \omega_{k}  \,\frac{\left(\omega_{k}^{2}+\omega_{0}^{2}\right)\tilde{f}\left(\omega_{0}\right)+2\omega_{k}^{2}\tilde{f}\left(\omega_{k}\right)}{\left(\omega_{k}^{2}-\omega_{0}^{2}\right)^{2}};
\end{equation}
this should be compared with Eq.~\eqref{eq:harmonicoscil}. Moreover, by taking the free-particle limit, one obtains:
\begin{equation} \label{eq:bzerofree}
\frac{d}{dk}\Gamma_k =\frac{1}{2} \left. \frac{d}{dk}\Gamma_k \right|_{\text{\tiny white}} \cdot\left[\frac{\tilde{f}\left(0\right)}{2}+\tilde{f}\left(\omega_{k}\right)\right],
\end{equation}
which differs from both Eq.~\eqref{eq:sdfsdl} and Eq.~\eqref{eq:fdgdfg} in the limit $\beta \rightarrow 0$. We have checked the result also by solving the problem from scratch, to first perturbative order~\cite{note}. 

From the mathematical point of view the reason for such a discrepancy is, once again, simple. In the exact calculation terms appear, which are proportional to $e^{-(\omega_0^2 \beta/2m)t}$ and therefore vanish in the large time limit. On the other hand, if one takes the limit $\beta \rightarrow 0$ before the large time limit, the decaying exponential is approximated by 1, and the associated terms contribute to the final result.
From the physical point of view, what happens is that, by stopping the computation at the first perturbative order in the electromagnetic constant $e$, one does not take into account all processes where virtual photons are emitted and reabsorbed. Actually, this is one way to look at the physical meaning of the exponential $e^{-(\omega_0^2 \beta/2m)t}$. 

The reason why the perturbative approach is problematic is that the interaction between the noise and the particle is different from the interactions usually studied (perturbatively) in quantum field theory. In our case the noise always interacts with the particle, so the typical scattering picture, where the particle is asymptotically free and the interaction is localized in a well defined region, is not true anymore. Accordingly, because the noise effects can build up in time, even if the interaction is very weak the perturbative approach fails when the large time limit $t \rightarrow \infty$ is taken. The solution found in~\cite{abd}---confining the noise in space and using final wave packets---reduces the interaction with the noise to a standard scattering process, for which perturbative techniques are valid. 

Our results here are in agreement with this picture. We have shown that for a bounded particle, no matter how weakly, when the emission rate is computed exactly the unphysical term is not present anymore, even without introducing any noise confinement. More precisely, we have seen that, when the emission rate is computed exactly, new contributions appear that, in the large time limit, suppress the term containing the unphysical factor.
Therefore the root of the problem lies in how the electromagnetic field reacts to the particle being accelerated by the noise. Constraining the effect of the noise by confining it in space/time allows to use the first order perturbation theory. In the other cases, also higher orders become relevant.

Our analysis has been carried out using the QMUPL model. It would be interesting to check if the same results are true also for the CSL model. However, for such a model, solving the Heisenberg equations for the operators, treating all the interactions exactly, is not possible. A first attempt might be to treat perturbatively both the electromagnetic and the noise interactions and carry the calculation up to the second order, instead of stopping at the first order. However, this poses a problem: the number of Feynman diagrams to compute is huge, as an easy inspection to the perturbative series shows. 
A more promising idea is to perform the computation by treating the electromagnetic interaction exactly and the noise interaction perturbatively. This is suggested by the fact that the exponential damping factor $e^{-(\omega_0^2 \beta/2m)t}$, the one that suppresses the term proportional to the unphysical factor, does not depend by $\lambda$ but only on $e$. This will be subject of future research.   
\section{Semiclassical derivation of the emission rate}
In this section we derive the emission rate for a free particle and for an harmonic oscillator using a semiclassical method. 
Such an approach has  been used also in appendix A of~\cite{ad1} for the CSL model and in~\cite{diosi} to compute the radiation emission due to the interaction with a stochastic gravitational background. The starting point is the Larmor formula:
\begin{equation} \label{eq:P}
P\left(t\right)=\frac{e^{2}}{6\pi\epsilon_{0}c^{3}}a^{2}\left(t\right),
\end{equation}
where $a\left(t\right)$ is the (modulus of the) acceleration of the charged particle and $P\left(t\right)$ is the total radiated power. 
Since the emission rate computed in the previous sections gives the number of photons emitted with energy $\omega_{k}=kc$, the total emitted power  is related to the emission rate by the relation:
\begin{equation} \label{eq:P-gamma}
P\left(t\right)=\int dk\hbar\omega_{k}\frac{d\Gamma\left(t\right)}{dk}=\int d\omega_{k}\frac{\hbar\omega_{k}}{c}\frac{d\Gamma\left(t\right)}{dk}.
\end{equation}
In order to find the emission rate starting from Eq.~\eqref{eq:P}, we rewrite it in terms of the Fourier transform of the acceleration:
\begin{equation} \label{eq:P2}
P\left(t\right)=\frac{e^{2}}{6\pi\epsilon_{0}c^{3}}\sum_{j=1}^{3}\frac{1}{\left(2\pi\right)^{2}}\int d\nu d\omega e^{i\left(\nu+\omega\right)t}a_{j}\left(\nu\right)a_{j}\left(\omega\right),
\end{equation}
where $a_j$ is the $j$-th cartesian component of the acceleration. This formula, together with the relation
\begin{equation} \label{eq:corrfourier}
\mathbb{E}\left[w_{j}\left(\nu\right)w_{j'}\left(\omega\right)\right]=\delta_{j,j'}4\pi\delta\left(\omega+\nu\right)\tilde{f}\left(\omega\right),
\end{equation}
where $w_{j}\left(\nu\right)$ is the Fourier transform of the noise $w_{j}\left(t\right)$ and $\tilde{f}\left(\omega\right)$ is defined in Eq.~\eqref{eq:ftilde}, is all we need to find the emission rate.

\vspace{.2cm}
\noindent {\it Free particle}.
For a free particle under the influence of the noise, acceleration is given by:
\begin{equation} \label{eq:afree}
\mathbf{a}\left(t\right)=\frac{\sqrt{\lambda}\hbar}{m}\mathbf{w}\left(t\right).
\end{equation}
Note that $\hbar$ appears in this classical calculation simply in order to keep the strength of the noise the same as in the quantum case. Inserting Eq.~\eqref{eq:afree} in Eq.~\eqref{eq:P2} and taking the noise average, one gets:
\begin{equation} \label{eq:Pfree}
P\left(t\right)=\int d\omega\frac{e^{2}\lambda\hbar^{2}}{2\pi^{2}\epsilon_{0}m^{2}c^{3}}\tilde{f}\left(\omega\right),
\end{equation}
where we used the relation given in Eq.~\eqref{eq:corrfourier}. By a comparison between this equation and Eq.~\eqref{eq:P-gamma} the emission rate is:
\begin{equation} \label{eq:ratefreesemi}
\frac{d\Gamma\left(t\right)}{dk}=\frac{e^{2}\lambda\hbar}{2\pi^{2}\epsilon_{0}m^{2}c^{2}\omega_{k}}\tilde{f}\left(\omega_{k}\right),
\end{equation}
that it is exactly the emission rate given in Eq.~\eqref{eq:sdfsdl} without the unphysical factor $\tilde{f}\left(0\right)$. It is remarkable that using this semiclassical method the factor $\tilde{f}\left(0\right)$ does not appear. The reason for this behavior is not clear and will be subject to future research.

\vspace{.2cm}
\noindent {\it Harmonic oscillator.}
The acceleration of a particle bounded by an harmonic potential and subject to a noise is:
\begin{eqnarray} \label{eq:aharmonic}
a_{j}\!\left(t\right) & = & \frac{\sqrt{\lambda}\hbar}{m}\!\left[w_{j}\!\left(t\right)+\omega_{0}\cos\left(\omega_{0}t\right)f_{s}\!\left(t\right)-\omega_{0}\sin\left(\omega_{0}t\right)f_{c}\!\left(t\right)\right] \nonumber \\
& + & \text{terms not depending on the noise,}
\end{eqnarray}
where we have introduced $f_{s}\left(t\right):=\int_{0}^{t}dx\sin\left(\omega_{0}x\right)w_{j}\left(x\right)$ and $f_{c}\left(t\right):=\int_{0}^{t}dx\cos\left(\omega_{0}x\right)w_{j}\left(x\right)$. 
By using the convolution theorem, we can write the Fourier transform of the second and the third terms as follows:  
\begin{equation} \label{eq:FTcos}
F_{\omega}\left\{ \cos\left(\omega_{0}t\right)f_{s}\left(t\right)\right\} =\frac{1}{2}F_{\omega-\omega_{0}}\left(f_{s}\left(t\right)\right)+\frac{1}{2}F_{\omega+\omega_{0}}\left(f_{s}\left(t\right)\right),
\end{equation}
\begin{equation} \label{eq:FTsen}
F_{\omega}\left\{ \sin\left(\omega_{0}t\right)f_{c}\left(t\right)\right\} =\frac{i}{2}F_{\omega+\omega_{0}}\left(f_{c}\left(t\right)\right)-\frac{i}{2}F_{\omega-\omega_{0}}\left(f_{c}\left(t\right)\right),
\end{equation}
where $F_{\omega}\left\{f\left(t\right)\right\}$ denotes the Fourier transform of $f\left(t\right)$.
The Fourier transform of the acceleration becomes:
\begin{eqnarray}
a_{j}\left(\omega\right) & = & \frac{\sqrt{\lambda}\hbar}{m}\left\{ w_{j}\left(\omega\right)+i\frac{\omega_{0}}{2}\left[F_{\omega-\omega_{0}}\left(\int_{0}^{t}dxe^{-i\omega_{0}x}w_{j}\left(x\right)\right)\right.\right. \label{eq:FTaharmonic}\nonumber \\
 &-& \left.\left. F_{\omega+\omega_{0}}\left(\int_{0}^{t}dxe^{i\omega_{0}x}w_{j}\left(x\right)\right)\right]\right\} \nonumber \\
 & + & \text{terms not depending on the noise.}
\end{eqnarray}
The two Fourier transforms in the above equation can be computed with an integrating by parts:
\begin{eqnarray}
F_{\omega-\omega_{0}}\left(\int_{0}^{t}dxe^{-i\omega_{0}x}w_{j}\left(x\right)\right)\; & = & \;\frac{w_{j}\left(\omega\right)}{i\left(\omega-\omega_{0}\right)}, \label{eq:FT-}\\
F_{\omega+\omega_{0}}\left(\int_{0}^{t}dxe^{i\omega_{0}x}w_{j}\left(x\right)\right) \;\;\;&=&\; \frac{w_{j}\left(\omega\right)}{i\left(\omega+\omega_{0}\right)}. \label{eq:FT+}
\end{eqnarray}
Here we have neglected the terms that oscillate infinitely fast with respect to $\omega$. This is justified by the fact that, in the formula for the total emitted power, an integral over $\omega$ appears, which is zero for these terms. Using Eq.~\eqref{eq:FT-} and Eq.~\eqref{eq:FT+} the expression for $a_{j}\left(\omega\right)$ becomes:
\begin{equation} \label{eq:FTaharmonic2}
a_{j}\left(\omega\right)=\frac{\sqrt{\lambda}\hbar}{m}w_{j}\left(\omega\right)\left[\frac{\omega^{2}}{\left(\omega^{2}-\omega_{0}^{2}\right)}\right].
\end{equation}
Using Eq.~\eqref{eq:P2} and taking the average over the noise we get:
\begin{equation}
P\left(t\right)=\frac{e^{2}\lambda\hbar^{2}}{2\pi^{2}\epsilon_{0}m^{2}c^{3}}\int d\omega\frac{\omega^{4}}{\left(\omega^{2}-\omega_{0}^{2}\right)^{2}}\tilde{f}\left(\omega\right),
\end{equation}
which means that the emission rate is:
\begin{equation}
\frac{d\Gamma\left(t\right)}{dk}=\frac{e^{2}\lambda\hbar}{2\pi^{2}\epsilon_{0}m^{2}c^{2}}\frac{\omega_{k}^{3}}{\left(\omega_{k}^{2}-\omega_{0}^{2}\right)^{2}}\tilde{f}\left(\omega_{k}\right).
\end{equation}  
This is equivalent to the quantum formula of Eq.~\eqref{eq:harmonicoscil}. 
 
\emph{Acknowledgements.}
Both authors wish to express their deep gratitude to S.L. Adler and D. D\"urr for many enjoyable and very stimulating discussions on this topic. They also wish to thank the anonymous referee who suggested to compute the rate using semiclassical methods. They both acknowledge partial financial support from INFN and COST (MP1006). A.B. acknowledges partial support also from NANOQUESTFIT and The
John Templeton Foundation project `Experimental and
theoretical exploration of fundamental limits of quantum
mechanics'.

\section*{EXPLICIT DERIVATION OF THE EMISSION RATES USING THE SOLUTIONS OF THE HEISENBERG EQUATIONS} 
This is the supplementary material for the article ``Spontaneous photon emission from a non-relativistic free charged particle in collapse models: a case study". In section VII we summarized the results found in the reference~[18]. In section VIII we show the derivation of the Eqs. (9), (15) and (17) of the letter.
\section{Exact solutions of the Heisenberg equations} \label{sec:uno}

In this section we summarize all the results which are relevant for our analysis and which can be found in reference~[18] of the letter. As show there, the non-linear dynamics of the QMUPL model can be linearized by taking an imaginary noise. This new dynamics, being linear, does not induce the collapse of the wave function. Nevertheless, it reproduces the same master equation as that associated to the non-linear equation. So, as far as we are only interested in computing the mean value of observables, like the emission rate, the linearized dynamics gives the same results as the non-linear one. The linearized dynamics is described by this stochastic Schr\"odinger equation:
\begin{equation} \label{eq:ghdfs}
i\hbar \frac{d}{dt} \psi_t = \left[ H - \sqrt{\lambda} \hbar\,
{\bf q}\cdot {\bf w}(t)  \right] \psi_t.
\end{equation}
The Hamiltonian $H$ is:
\begin{equation} \label{eq:H}
H \; = \; \frac{1}{2m_0}({\bf p} - e {\bf A})^2 + \frac{1}{2}
\kappa\, {\bf q}^2 + \frac{1}{2}\epsilon_0 \int d^3x \, \left[ {\bf
E}^2 + c^2 {\bf B}^2 \right].
\end{equation}
The meaning of all the symbols in Eq.~(\ref{eq:H}) is described in the letter.
The plane waves decomposition of the vector potential ${\bf A}$, in the dipole approximation, reads:
\begin{equation}
{\bf A}({\bf x}) \; := \; \sqrt{\frac{\hbar}{\epsilon_0}} \sum_{\mu}
\int d^3k\, \frac{g({\bf k})}{\sqrt{2\omega_k}}\,
\vec{\epsilon}_{\bf k \mu} \left[ a_{\bf k \mu}^{\phantom \dagger} +
a_{\bf k \mu}^{\dagger} \right];
\end{equation}
where $\omega_k = c k$ ($k = | {\bf k} |$) is the frequency
corresponding to the wave vector ${\bf k}$; $\vec{\epsilon}_{\bf k
\mu}$ ($\mu = 1,2$) are the linear polarization vectors and $a_{\bf
k \mu}^{\dagger}$, $a_{\bf k \mu}^{\phantom \dagger}$ are the
creation and annihilation operators satisfying the standard
commutation relations:
\begin{equation}
[a_{\bf k \mu}^{\phantom \dagger}, a_{\bf k' \mu'}^{\dagger}] \; =
\; \delta_{\mu \mu'}^{\phantom \dagger} \delta^{(3)}({\bf k} - {\bf
k}').
\end{equation}
The function $g({\bf k})$ is the Fourier transform of the charge distribution and it reduces to $1/(2\pi)^{3/2}$ for a point like particle.

The second term in the right hand side of Eq.~(\ref{eq:ghdfs}) is a random potential depending on the position ${\bf q}$ of the particle and on the noise  ${\bf w}(t)$. This noise has zero mean and a generic correlation given by $\mathbb{E} [ w_{i}(s) w_{j} (s')] = \delta_{ij} f(s-s')$, where the subscripts $i,j=1,2,3$ label the three directions in the space. This means that, on the contrary to what was done in reference~[18], here we consider the more generic situation of a (isotropic) colored noise. However, since the solutions of the Heisenberg equations for the operators are formally equivalent both in the white and in the colored noise cases, we are allowed to take the solutions found in reference~[18], without any further calculation.

The Heisenberg equations of motion for the position of the particle ${\bf q}(t)$, its conjugate momentum ${\bf p}(t)$ and the electromagnetic-field operators $a_{\bf k \mu}^{\dagger}(t)$ are:
\begin{eqnarray}
\frac{d {\bf p}}{dt} & = & - \kappa\, {\bf q} \; + \; \sqrt{\lambda}
\hbar {\bf w}(t), \label{eq:dfgsd}\\
& & \nonumber \\ \frac{d {\bf q}}{dt} & = & \frac{\bf p}{m_0} \; - \;
\frac{e}{m_0}\, {\bf A}, \label{eq:dfgsd2}
\end{eqnarray}
\begin{equation} \label{eq:dfgsd3}
\frac{d a_{\bf k \mu}^{\dagger}}{dt} = {\phantom -}i \omega_{\bf k}
a_{\bf k \mu}^{\dagger} - \frac{ie}{\sqrt{\hbar\epsilon_0} m_0}\,
\frac{g({\bf k})}{\sqrt{2\omega_k}} \vec{\epsilon}_{\bf k
\mu}^{\phantom \dagger} \cdot {\bf p} \; + \; \frac{i
e^2}{\epsilon_0 m_0} \, \frac{g({\bf k})}{\sqrt{2\omega_k}}
\vec{\epsilon}_{\bf k \mu}^{\phantom \dagger} \cdot \sum_{\mu'} \int
d^3 k' \frac{g({\bf k}')}{\sqrt{2\omega_{k'}}} \vec{\epsilon}_{\bf
k' \mu'}^{\phantom \dagger} \left[ a_{\bf k' \mu'}^{\phantom
\dagger} + a_{\bf k' \mu'}^{\dagger} \right];\;\;\;\;
\end{equation}
the equation for $a_{\bf k \mu}^{\phantom \dagger}(t)$ can be
obtained from the previous one by taking the hermitian conjugate.

By solving the above set of coupled linear differential equations, one obtains:
\begin{eqnarray}
{\bf q}(t) & = & \left[ 1 - \kappa\, F_1(t) \right] {\bf q}(0) \;
+ \; F_0(t)\, {\bf p}(0) \nonumber \\
& & - e\, \sqrt{\frac{\hbar}{\epsilon_0}} \, \sum_{\mu} \int d^3 k
\, \frac{g({\bf k})}{\sqrt{2\omega_k}} \, \vec{\epsilon}_{\bf k
\mu}^{\phantom \dagger} \left[ G^{+}_{1}(k,t)\, a_{\bf k
\mu}^{\phantom \dagger}(0) +
G^{-}_{1}(k,t)\, a_{\bf k \mu}^{\dagger}(0) \right] \nonumber \\
& & + \sqrt{\lambda}\hbar \int_0^t ds F_0(t - s) {\bf
w}(s), \label{eq:fsdgh} \\ & & \nonumber\\
{\bf p}(t) & = & - \kappa \left[ t - \kappa\, F_2(t) \right] {\bf
q}(0) \; + \; \left[ 1 - \kappa\, F_1(t) \right] {\bf p}(0)
\nonumber \\
& & + \kappa\, e \, \sqrt{\frac{\hbar}{\epsilon_0}} \, \sum_{\mu}
\int d^3 k \, \frac{g({\bf k})}{\sqrt{2\omega_k}} \,
\vec{\epsilon}_{\bf k \mu}^{\phantom \dagger} \left[
G^{+}_{0}(k,t)\, a_{\bf k \mu}^{\phantom \dagger}(0) +
G^{-}_{0}(k,t)\, a_{\bf k \mu}^{\dagger}(0) \right] \nonumber \\
& & + \sqrt{\lambda} \hbar \int_0^t ds \left[ 1 - \kappa\,
F_1(t-s) \right] {\bf w}(s) \\ & & \nonumber \\
a_{\bf k \mu}^{\dagger}(t) & = & e^{i \omega_k t} a_{\bf k
\mu}^{\dagger}(0) \; - \; \frac{ie}{\sqrt{\hbar\epsilon_0}}\,
\frac{g({\bf k})}{\sqrt{2\omega_k}} \, \vec{\epsilon}_{\bf k
\mu}^{\phantom \dagger} \cdot \left[ G^{-}_{1}(k,t)\, {\bf p}(0) -
\kappa \, G^{-}_{0}(k,t)
\, {\bf q}(0) \right] \nonumber \\
& & + \frac{i e^2}{\epsilon_0} \, \frac{g({\bf
k})}{\sqrt{2\omega_k}} \, \vec{\epsilon}_{\bf k \mu}^{\phantom
\dagger} \cdot \sum_{\mu'} \int d^3 k' \, \frac{g({\bf
k}')}{\sqrt{2\omega_{k'}}} \, \vec{\epsilon}_{\bf k' \mu'}^{\phantom
\dagger} \left[ G^{-}_{+}(k,k',t)\, a_{\bf k' \mu'}^{\phantom
\dagger}(0) +
G^{-}_{-}(k,k',t)\, a_{\bf k' \mu'}^{\dagger}(0) \right] \nonumber \\
& & - ie\sqrt{\frac{\hbar \lambda}{\epsilon_0}} \, \frac{g({\bf
k})}{\sqrt{2\omega_k}} \, \vec{\epsilon}_{\bf k \mu}^{\phantom
\dagger} \cdot \int_0^t ds\,
G^{-}_{1}(k,t-s) {\bf w}(s), \label{eq:fed} \\ & & \nonumber \\
a_{\bf k \mu}^{\phantom \dagger}(t) & = & e^{-i \omega_k t} a_{\bf k
\mu}^{\phantom \dagger}(0) \; + \;
\frac{ie}{\sqrt{\hbar\epsilon_0}}\, \frac{g({\bf
k})}{\sqrt{2\omega_k}} \, \vec{\epsilon}_{\bf k \mu}^{\phantom
\dagger} \cdot \left[ G^{+}_{1}(k,t)\, {\bf p}(0) - \kappa \,
G^{+}_{0}(k,t)
\, {\bf q}(0) \right] \nonumber \\
& & - \frac{i e^2}{\epsilon_0} \, \frac{g({\bf
k})}{\sqrt{2\omega_k}} \, \vec{\epsilon}_{\bf k \mu}^{\phantom
\dagger} \cdot \sum_{\mu'} \int d^3 k' \, \frac{g({\bf
k}')}{\sqrt{2\omega_{k'}}} \, \vec{\epsilon}_{\bf k' \mu'}^{\phantom
\dagger} \left[ G^{+}_{+}(k,k',t)\, a_{\bf k' \mu'}^{\phantom
\dagger}(0) +
G^{+}_{-}(k,k',t)\, a_{\bf k' \mu'}^{\dagger}(0) \right] \nonumber \\
& & + ie\sqrt{\frac{\hbar \lambda}{\epsilon_0}} \, \frac{g({\bf
k})}{\sqrt{2\omega_k}} \, \vec{\epsilon}_{\bf k \mu}^{\phantom
\dagger} \cdot \int_0^t ds\, G^{+}_{1}(k,t-s) {\bf w}(s),
\label{eq:fe}
\end{eqnarray}
In the previous formulas, we have introduced the following
functions:
\begin{eqnarray}
F_n(t) & := & \int_{\Gamma} \frac{dz}{2\pi i} \, \frac{e^{zt}}{z^n
H(z)}, \qquad\qquad\qquad\qquad\;\;\; n = 0,1,2, \label{eq:F}\\
G^{\pm}_n(k,t) & := & \int_{\Gamma} \frac{dz}{2\pi i} \, \frac{z^n
e^{zt}}{(z \pm i \omega_k) H(z)}, \qquad\qquad\qquad n = 0,1, \label{eq:G} \\
G^{\pm}_{\pm}(k,k't) & := & \int_{\Gamma} \frac{dz}{2\pi i} \,
\frac{z^2 e^{zt}}{(z \pm i \omega_k)(z \pm i \omega_{k'}) H(z)};
\label{eq:Gpm}
\end{eqnarray}
in the third expression, the upper $\pm$ refers to the first parenthesis, while
the lower one refers to the second parenthesis. In all the above formulas,
according to the theory of Laplace transform, the contour $\Gamma$ must be a
line parallel to the imaginary axis, lying to the right of all singularities of
the integrand.
The function $H(z)$, after the renormalization procedure, is:
\begin{equation} \label{eq:Hfun}
H(z) \; = \; \kappa + z^2 \left[ m - \beta \, z \right],
\qquad\qquad \beta \; = \; \frac{e^2}{6 \pi \epsilon_0 c^3} \;
\simeq \; 5.71 \times 10^{-54}\, \makebox{Kg s}.
\end{equation}
Since $F_n(t)$, $G^{\pm}_n(k,t)$  and $G^{\pm}_{\pm}(k,k't)$ are defined by contour integrals of functions containing $H(z)$ at denominator, it is fundamental to know the zeros of $H(z)$. Their approximate values are (see Appendix A in reference~[18]):
\begin{equation} \label{eq:sol}
z_1 \; \simeq \; \frac{m}{\beta} \; + \;  o(\omega_0), \quad \qquad
z_{2,3} \; \simeq \; - \frac{\omega_0^2 \beta}{2m} \, \pm \, i
\omega_0 \; + \; o(\omega_0^3).
\end{equation}
In the following we will need the functions $F_n(t)$ and $G^{\pm}_n(k,t)$ for $n=0,1$. Let us write such functions explicitly:
\begin{eqnarray}
F_{0}\left(t\right) & = & \sum_{l=1}^{3}e^{z_{l}t}\left[\frac{z-z_{l}}{H\left(z\right)}\right]_{z=z_{l}}= \label{eq:F0}\\ 
& = & -\frac{e^{z_{1}t}}{\beta\left(z_{1}-z_{2}\right)\left(z_{1}-z_{3}\right)}+\frac{e^{z_{2}t}}{\beta\left(z_{1}-z_{2}\right)\left(z_{2}-z_{3}\right)}-\frac{e^{z_{3}t}}{\beta\left(z_{1}-z_{3}\right)\left(z_{2}-z_{3}\right)},\nonumber\\
&&\nonumber\\
F_{1}\left(t\right) & = & \sum_{l=1}^{3}\frac{e^{z_{l}t}}{z_{l}}\left[\frac{z-z_{l}}{H\left(z\right)}\right]_{z=z_{l}}+\frac{1}{\beta z_{1}z_{2}z_{3}}= \label{eq:F1}\\ 
& = & -\frac{e^{z_{1}t}}{\beta z_{1}\left(z_{1}-z_{2}\right)\left(z_{1}-z_{3}\right)}+\frac{e^{z_{2}t}}{\beta z_{2}\left(z_{1}-z_{2}\right)\left(z_{2}-z_{3}\right)}-\frac{e^{z_{3}t}}{\beta z_{3}\left(z_{1}-z_{3}\right)\left(z_{2}-z_{3}\right)}+\frac{1}{\beta z_{1}z_{2}z_{3}},\nonumber\\
&&\nonumber\\
G_{0}^{\pm}\left(k,t\right) & = &  \sum_{l=1}^{3}\frac{e^{z_{l}t}}{\left(z_{l}\pm i\omega_{k}\right)}\left[\frac{z-z_{l}}{H\left(z\right)}\right]_{z=z_{l}}\mp\frac{e^{\mp i\omega_{k}t}}{H\left(\mp i\omega_{k}\right)}= \label{eq:G0}\\ 
& = & -\frac{e^{z_{1}t}}{\beta\left(z_{1}-z_{2}\right)\left(z_{1}-z_{3}\right)\left(z_{1}\pm i\omega_{k}\right)}+\frac{e^{z_{2}t}}{\beta\left(z_{1}-z_{2}\right)\left(z_{2}-z_{3}\right)\left(z_{2}\pm i\omega_{k}\right)}\nonumber\\
&  & -\frac{e^{z_{3}t}}{\beta\left(z_{1}-z_{3}\right)\left(z_{2}-z_{3}\right)\left(z_{3}\pm i\omega_{k}\right)}\mp\frac{e^{\mp i\omega_{k}t}}{\beta\left(z_{1}\pm i\omega_{k}\right)\left(z_{2}\pm i\omega_{k}\right)\left(z_{3}\pm i\omega_{k}\right)},\nonumber\\
&&\nonumber\\
G_{1}^{\pm}\left(k,t\right) & = &  \sum_{l=1}^{3}\frac{z_{l}e^{z_{l}t}}{\left(z_{l}\pm i\omega_{k}\right)}\left[\frac{z-z_{l}}{H\left(z\right)}\right]_{z=z_{l}}\mp\frac{i\omega_{k}e^{\mp i\omega_{k}t}}{H\left(\mp i\omega_{k}\right)}= \label{eq:G1}\\ 
& = & -\frac{z_{1}e^{z_{1}t}}{\beta\left(z_{1}-z_{2}\right)\left(z_{1}-z_{3}\right)\left(z_{1}\pm i\omega_{k}\right)}+\frac{z_{2}e^{z_{2}t}}{\beta\left(z_{1}-z_{2}\right)\left(z_{2}-z_{3}\right)\left(z_{2}\pm i\omega_{k}\right)}\nonumber\\
&  & -\frac{z_{3}e^{z_{3}t}}{\beta\left(z_{1}-z_{3}\right)\left(z_{2}-z_{3}\right)\left(z_{3}\pm i\omega_{k}\right)}\mp\frac{i\omega_{k}e^{\mp i\omega_{k}t}}{\beta\left(z_{1}\pm i\omega_{k}\right)\left(z_{2}\pm i\omega_{k}\right)\left(z_{3}\pm i\omega_{k}\right)}.\nonumber\\
&&\nonumber
\end{eqnarray}
All these functions contain terms that depend on time in four different ways. There are terms containing an exponential with $z_1$. These are runaway terms, which we pragmatically ignore. There are constant and oscillating terms that can give important contributions to the rate. Finally there are terms containing exponentials with $z_2$ or $z_3$: this terms are crucial for our analysis. Indeed, when $\beta$ and $\omega_0$ are different from zero, these terms decay and they do not contribute to the rate for large times. However, when $\omega_0=0$ (free particle limit) or when $\beta=0$ (interaction with the electromagnetic field treated at the first perturbative order), these terms do not decay anymore and they contribute to the final rate, giving rise to an unphysical term (see further comments in the letter).

\section{Explicit derivation of Eqs. (9), (15) and (17)}

In this section we show how to compute the emission rate for a free particle (Eq.(9) of the letter) and for a particle bounded by an harmonic oscillator potential (Eq.(15) of the letter). We also show why, when doing the computation by treating the interaction with the electromagnetic field at the lowest order, for a particle in an harmonic oscillator potential one finds Eq.(17) of the letter instead of the correct result given by Eq.(15). 

The emission rate is defined as:
\begin{equation} \label{eq:ratedef}
\frac{d}{dt}\Gamma_{k} =8\pi k^2 \frac{d}{dt} {\mathbb E} [\langle \phi | a_{\bf k \mu}^{\dagger}(t) a_{\bf k \mu}(t) | \phi \rangle].
\end{equation}
Looking at the Eq.~\eqref{eq:fed} for $a_{\bf k \mu}^{\dagger}(t)$ and Eq.~\eqref{eq:fe} for $a_{\bf k \mu}(t)$, we see that the only non zero contribution due to the noise, the one proportional to $\lambda$, is given by the product of the last terms. The reason why the terms proportional to $\sqrt{\lambda}$, that contain only one noise, do not contribute is that the noise has zero average. Recalling that the noise correlation is $\mathbb{E} [ w_{i}(s) w_{j} (s')] = \delta_{ij} f(s-s')$, it is easy to get, for a point like particle ($g({\bf k})=1/(2\pi)^{3/2}$): 
\begin{equation} \label{eq:rate}
\frac{d}{dt}\Gamma_{k} = \frac{\lambda\hbar e^{2}}{2\pi^{2}\varepsilon_{0}c^{2}}\omega_{k}\frac{d}{dt}\int_{0}^{t}ds\int_{0}^{t}ds'G_{1}^{-}\left(k,t-s\right){G}_{1}^{+}\left(k,t-s'\right)f\left(s-s'\right).
\end{equation}
This is equivalent to the formula (4) of the letter. Writing more explicitly the functions $G^{\pm}_1(k,t)$ given in Eq.~\eqref{eq:G1} one gets (the runaway terms containing $e^{{z_1}t}$ are neglected):
\begin{eqnarray}\label{eq:ratefineteT}
&&\frac{d}{dk}\Gamma_k=\frac{\lambda\hbar e^{2}}{2\pi^{2}\varepsilon_{0}c^2}\omega_{k}\;\cdot \\
& \cdot & \left\{\sum_{\ell,\ell'=2}^{3}\frac{z_{\ell}z_{\ell'}}{\left(z_{\ell}-i\omega_{k}\right)\left(z_{\ell'}+i\omega_{k}\right)}\left[\frac{1}{\beta}\prod_{\underset{j\neq \ell}{j=1}}^{3}\frac{1}{z_{\ell}-z_{j}}\right]\left[\frac{1}{\beta}\prod_{\underset{j\neq \ell'}{j=1}}^{3}\frac{1}{z_{\ell'}-z_{j}}\right]e^{\left(z_{\ell}+z_{\ell'}\right)t} [F_t(-z_{\ell'}) + F_t(-z_{\ell})]\right.\;\;\;\; \nonumber \\
& & 
-i\sum_{\ell=2}^{3}\frac{\omega_{k}z_{\ell}}{H\left(-i\omega_{k}\right)\left(z_{\ell}-i\omega_{k}\right)}\left[\frac{1}{\beta}\prod_{\underset{j\neq \ell}{j=1}}^{3}\frac{1}{z_{\ell}-z_{j}}\right]e^{\left(z_{\ell}-i\omega_{k}\right)t} [ F_t(i\omega_k) + F_t(-z_{\ell})] + \nonumber \\
& & 
+i\sum_{\ell=2}^{3}\frac{\omega_{k}z_{\ell}}{H\left(+i\omega_{k}\right)\left(z_{\ell}+i\omega_{k}\right)}\left[\frac{1}{\beta}\prod_{\underset{j\neq \ell}{j=1}}^{3}\frac{1}{z_{\ell}-z_{j}}\right]e^{\left(z_{\ell}+i\omega_{k}\right)t} [ F_t(-z_{\ell}) + F_t(-i\omega_k)] + \nonumber \\
& & 
\left.+\frac{\omega_{k}^{2}}{H\left(+i\omega_{k}\right)H\left(-i\omega_{k}\right)} [ F_t(i\omega_k) + F_t(-i\omega_k) ]\right\},\nonumber 
\end{eqnarray}
where $F_t(z)$ is defined as:
\begin{equation}
F_t(\alpha) \; = \; \int_0^t dx f(x) e^{\alpha x}.
\end{equation}

The large time limit is easy to understand. Since $z_{2,3}  =  - \frac{\omega_0^2 \beta}{2m} \, \pm \, i
\omega_0$, the first three lines contain terms that vanish exponentially. Then the only term that survives is the one in the fourth line, which, for $t \rightarrow \infty$, gives:
\begin{equation} \label{eq:correctsp}
\frac{d}{dk}\Gamma_k = \frac{\lambda\hbar e^{2}}{2\pi^{2}\varepsilon_{0}c^2} \frac{\omega_{k}^3}{m^{2}\left(\omega_{0}^{2}-\omega_{k}^{2}\right)^{2}+\beta^{2}\omega_{k}^{6}}\tilde{f}\left(\omega_{k}\right),
\end{equation}
where $\tilde{f}(k) = \int_{- \infty}^{+ \infty} f(s) e^{iks} ds$. This is exactly Eq. (15) of the letter. 

Let us study the free particle case, which correspond to $\omega_{0}=0$. This implies that $z_{2,3} = 0$. In such a case the third and the fourth line of Eq.~\eqref{eq:ratefineteT} give contributions that oscillate in time, due to the presence of $e^{{\pm} i\omega_{k} t}$. In the large time limit such contributions average to zero. However, the same is not true for the term in the second line: this term now is constant and it is proportional to $F_t(0)$. In the large time limit the rate become:
\begin{equation} \label{eq:freesp}
\frac{d}{dk}\Gamma_k= \frac{1}{2} \left. \frac{d}{dk}\Gamma_k \right|_{\text{\tiny white}} \cdot \left[\tilde{f}\left(0\right)+\frac{m^2}{m^{2}+\beta^{2}\omega_{k}^{2}}\tilde{f}\left(\omega_{k}\right)\right],
\end{equation}
 that is exactly Eq. (9) of the letter.

To conclude this section we study what happens if one stops the computation at the lowest order in the electromagnetic constant, that is the order $e^2$. Since there is a factor $e^2$ in front of all the terms in Eq.~\eqref{eq:ratefineteT} and since all other $e$'s are contained in $\beta=\frac{e^2}{6 \pi \epsilon_0 c^3}$, performing the computation at the order $e^2$ is equivalent to taking the solution of Eq.~\eqref{eq:ratefineteT} and set $\beta=0$. As consequence, in this case we have $z_{2,3}= \pm \, i\omega_0$. Since the real negative part of $z_{2,3}$, responsible of the decay behavior, has reduced to zero, once again the decay terms become oscillating (second and third line) and constant (first line). In particular, the first line is proportional to $F_t(\omega_{0})$. It is straightforward to see that, in the large time limit $t\rightarrow\infty$, one gets:
\begin{equation} \label{eq:bzerosp}
\frac{d}{dk}\Gamma_k = \frac{e^{2}\hbar\lambda}{4\pi^{2}\varepsilon_{0}c^{2}m^{2}}\, \omega_{k}  \,\frac{\left(\omega_{k}^{2}+\omega_{0}^{2}\right)\tilde{f}\left(\omega_{0}\right)+2\omega_{k}^{2}\tilde{f}\left(\omega_{k}\right)}{\left(\omega_{k}^{2}-\omega_{0}^{2}\right)^{2}},
\end{equation}
that is the Eq. (17) of the letter. If we compare this equation with the correct result given by Eq.~\eqref{eq:correctsp}, in the limit $\beta\rightarrow0$, we see that there is an extra term containing $\tilde{f}\left(\omega_{0}\right)$. Therefore we conclude that a computation of the rate, in which the electromagnetic interaction is analyzed at the lowest order in $e$, is not reliable.


\begin{thebibliography}{1}
\bibitem{com2} In this way we also correct a typo in Eq.~(41) of~\cite{bd}.
\end{thebibliography}
\end{document}